\documentstyle[preprint,revtex]{aps}
\tightenlines

\begin{document}
\draft
\begin{title}
Quark-based Description of Nuclear Matter \\
with Simulated Annealing
\end{title}
\author{George M. Frichter}
\begin{instit}
Department of Physics, \\
Florida State University, Tallahassee, FL 32306
\end{instit}
\author{J. Piekarewicz}
\begin{instit}
Supercomputer Computations Research Institute, \\
Florida State University, Tallahassee, FL 32306
\end{instit}

\begin{abstract}
We calculate ground-state properties of a many-quark system in
the string-flip model using variational Monte Carlo methods.
The many-body potential energy of the system is determined by
finding the optimal grouping of quarks into hadrons. This (optimal)
assignment problem is solved by using the stochastic optimization
technique of simulated annealing. Results are presented for the
energy and length-scale for confinement as a function of density.
These results show how quarks clustering decreases with density and
characterize the nuclear- to quark-matter transition. We compare our
results to a previously published work with a similar model which
uses, instead, a pairing approach to the optimization problem.
\end{abstract}
\clearpage

\narrowtext
\section{Introduction}
\label{intro}

Hadronic models have been very successful in describing the
interactions which give rise to the properties of nuclei and
nuclear matter. However, at a more fundamental level, our current
understanding of the strong force gives us a picture of fermion
fields (quarks) interacting via gauge fields (gluons) generated by an
underlying local SU(3)$_{\rm c}$ symmetry. Perhaps the greatest challenge
in the field of nuclear structure is to understand the emergence of
nuclear properties from the underlying QCD dynamics in which quark and
gluon fields are the fundamental degrees of freedom. The success of
experimental efforts at CEBAF and RHIC aimed at identifying signatures
of the quark substructure in nuclei will rely on the ability of
theoretical models to identify the appropriate observables. One might
be interested, for example, in understanding how are hadronic
properties such as the nucleon form factor modified by its interactions
with the medium? Are there new modes of excitation in many-quark systems
({\it e.g.}, coherent oscillations in density, spin,
flavor, or color) not present in purely hadronic models? Also, what is the
nature of the nuclear-matter to quark-matter transition? The hope is that
simple models such as the one to be discussed here will provide a first
step towards developing ``realistic'' many-quark models which identify
the relevant nuclear observables.

In this paper we will be working with a string-flip model which has been
described in detail elsewhere~\cite{j2,h1,j3,j1}. For the many-quark
system, the model attempts to reproduce nuclear/quark-matter properties
using exclusively quark degrees of freedom. Confining strings which
adjust instantaneously to the changing constituent quark positions
(adiabatic approximation) are used as a minimum representation of the
extremely complex quark-gluon dynamics. In this model, there are no strings
connecting quarks ``belonging'' to different hadrons. Hence, the force
saturates,
thus allowing clusters to separate, and avoids the emergence of long-range van
der Waals forces. Among the desirable properties of this potential are explicit
quark-exchange symmetry, asymptotic freedom and confinement.
However, because the model is a many-body generalization of the
nonrelativistic constituent quark model it is not chirally symmetric,
explicitly gauge invariant, nor is it Lorentz invariant.

A critical feature shared by all models of this type is the need to
determine an optimal grouping of the quarks which minimizes the total
potential energy stored in the strings. This assignment problem is
similar in nature to the well known traveling salesman problem in the
field of combinatorial optimization. Hence,
except for simulations with a very small number of quarks, the
computational cost of an exhaustive search over the possible string
configurations is beyond the capabilities of even the most powerful
computers available today. As a result, we present a model which uses the
technique of simulated annealing to solve the optimal grouping problem.
For the purposes of comparison, we will also describe a similar model which
uses an efficient pairing algorithm to solve a simpler assignment
problem \cite{j1}.

The paper is organized as follows. Section~\ref{sfm} gives a general
description of the string-flip model. In this section the many-body
potential is introduced as well as the techniques used in solving
the optimization problem. In Sec.~\ref{wavefun} we define the
one-parameter variational wavefunction that is employed to generate,
via Metropolis Monte Carlo, the quark configurations. Results
are presented in Sec.~\ref{res} while Sec.~\ref{concl} summarizes
the main results and points toward possible future work in this area.

\section{String-Flip Model}
\label{sfm}

\subsection{The Many-Body Potential}
\label{mbp}

In the present version of the string-flip model the quarks are grouped into
colorless triplets (red+green+blue=white) by harmonic confining
strings~\cite{j1}.
Governing the dynamics of $N=3A$ quarks grouped into $A$ colorless
hadrons is the following Hamiltonian,
\begin{equation}
{\bf H}=\sum_{i=1}^{N}-\frac{\nabla_i^2}{2M}+{\bf V}\:,
\end{equation}
where
\begin{equation}
{\bf V}={\rm Min}\sum_{i=1}^{A}\frac{k}{2}
[{(x_{i}^R-x_{i}^G)}^2+{(x_{i}^G-x_{i}^B)}^2
+{(x_{i}^B-x_{i}^R)}^2]\:.
\label{pot}
\end{equation}
Here $x_{i}^R$, $x_{i}^G$, and $x_{i}^B$ are the positions of
the red, green, and blue quarks belonging to hadron $i$ and
the minimization is taken over a (potentially enormous) set of
$(A!)^{2}$ possible string configurations. Furthermore, since
the quark mass $(M)$ and the spring constant $(k)$ are the only
dimensionfull quantities in the problem one is free to set
\begin{equation}
k=M=1\:,
\end{equation}
and, thus, measure all energies in units of the oscillator frequency
$\sqrt{k/M}$, and all lengths according to the oscillator length
$(Mk)^{-\frac{1}{4}}$. It is then easy to rescale any results according
to a particular choice of $k$ and $M$.

The potential energy for an isolated 3-quark cluster (nucleon) is,
in principle, an arbitrary confining function of the three quark
coordinates. In the simplest version of the model used here, however,
quarks are assumed to be confined by harmonic strings. For this case
a triangular (or $\Delta$) configuration yields identical results to the
$Y$ configuration which has quarks connected to a three-quark junction
as seems to be preferred by the underlying gauge symmetry of QCD (see
Fig.~\ref{3q01}). We have chosen the $\Delta$ configuration since it
simplifies the evaluation of the potential energy by eliminating
the calculation of a center-of-mass coordinate for each hadron.

In this model there are no long range van der Waals forces between hadrons.
The only (residual) interactions among hadrons are generated by the
possibility of quark exchange (change in grouping)
and the Pauli exclusion principle between quarks. Thus, the residual
interaction
between hadrons is short range (of the order of the root-mean-square radius
of the hadron). Furthermore, the potential is a truly many-body operator
({\it i.e.}, it cannot be reduced to a sum over two-body operators);
moving a single quark may cause all the quark groupings to change.

\subsection{The Three-quark Assignment Problem}
\label{tap}

For the many-quark system,
local $SU(3)_{\rm c}$ gauge symmetry demands that a color flux tube
(string) leaving one quark must connect to an anti-quark or to a three-quark
junction. For harmonic strings one can avoid the use of the Y-coupling
scheme and can connect the quarks directly to one
another. So, in our model, a string leaving one quark will connect to another.
The fundamental question is which one? Lattice QCD solves this assignment
problem automatically by evolving gluonic degrees of freedom according to
extremely complex dynamics resulting in various flux tube arrangements.
Unfortunately, the computational cost of solving full QCD on a lattice is
enormous. Perhaps for the forseeable future approximate models of the complex
QCD dynamics will be needed to give insight into the role of hadronic
substructure in nuclear physics. In the string-flip model used here the
requirement is to determine the string topology which minimizes the potential
energy. The flux tubes are assumed to adjust instantaneously to the changing
constituent quark positions (adiabatic approximation) and are used as a
minimum representation of the extremely complex quark-gluon dynamics.

For the problem of grouping $N=3A$ quarks into $A$ colorless hadrons, an
exhaustive search for the minimal configuration has a computational cost
proportional to $A!^2$. This becomes prohibitive for more than about 5
hadrons (note, for ten hadrons there are $10^{13}$ possible groupings!).
Another approach is obviously needed.

\subsubsection{Simulated Annealing}
\label{sa}

	To date, no efficient (power-law) algorithm has been developed
to solve the three-quark assignment problem. Thus, in order to
evaluate the many-quark potential [Eq.~(\ref{pot})] we resort
to simulated annealing.

Simulated annealing is a stochastic optimization technique which has been
successfully applied to problems of large dimension. For background on the
annealing approach to problems of combinatorial optimization, we refer the
reader to the excellent book by Otten and van Ginneken~\cite{Otten}.
In our problem the quantity to be minimized is the potential energy and we
choose the search space to be all colorless triplets of quarks. Here,
$N=3A$, and there are $A!^2$ possible string configurations to search over.

The annealing simulation proceeds much like the familiar Metropolis random
walk used in numerical integration except for the addition of a control
parameter, or temperature, which is initially set very high. A high
temperature is one which is much greater than typical energy differences
encountered by the random walker when taking steps. This means that the
probability for accepting the next step is nearly unity and the walker
wanders randomly throughout the search space, uninhibited by the structure
of the energy landscape. The control parameter is then slowly (adiabatically)
lowered so that the walker progressively visits lower energy regions with
increasing probability. By `slowly' we mean that the system must always
remain near equilibrium (quasi-equilibrium). Eventually, as the temperature
is taken lower and lower, the walker `freezes' into what is hopefully the
lowest possible energy configuration. The probability that the random walker
`freezes' into this global minimum instead of one of the innumerable local
minima is directly related to how well the criterion of quasi-equilibrium
was maintained throughout the cooling process and this in turn
is related to the amount of computation time invested. So, in practice, one
always solves the problem with some confidence level determined by
computational constraints.
For example, in the string-flip model, the estimate of ground-state
observables will be done via Metropolis Monte Carlo. Thus, the annealing
must be performed at each step of the Monte Carlo simulation of the quark
coordinates. This means an enormous number of annealing runs must be done
during a typical simulation. Obviously there will be limits on how long one
can spend on each annealing process.

The simple annealing algorithm used in this work is described in
Fig.~\ref{3qalg}.
By far the most important issue in developing this kind of simulation is the
selection of a cooling schedule and choosing the number of thermalization
steps.
In general, the number of thermalization steps could be a function of
temperature. In this particular simulation, we have chosen a constant number
of thermalization steps. The initial temperature should be chosen high
enough so that the entire configuration space is sampled but not so high that
computation time is wasted. Also, the cooling must be slow enough that the
random walker doesn't prematurely freeze into a local energy minimum
but not so slow that computation time is unnecessarily long.

Implementation of the annealing algorithm within the string-flip model
is complicated by the fact that it is imbedded in a larger Monte Carlo
simulation ({\it i.e.,} calculation of the expectation values
$\langle{\bf V}\rangle$)
of the quark coordinates. Each time the configuration
of quarks is modified, the energy landscape for the subsequent annealing
process changes slightly and it is the
structure of the energy landscape which drives the choice of cooling
parameters.
The energy landscape will also vary
significantly as a function of density. For example, in the
low density hadronic phase where the
quarks are tightly clustered, there is a large energy contrast between states
in the vicinity of the global minimum and states which lie farther away in the
configuration space. Now, as the density increases, the system experiences a
transition to a phase characterized by somewhat larger quark clusters. At
these higher densities, the energy difference between random groupings and
those near the global minimum is less than in the low density phase.
We can see then that a cooling schedule should be robust enough to perform
adequately at all densities of interest. We chose
our cooling parameters based on simulations in the high density region where
the annealing is somewhat more demanding.

Inevitably, a lot of numerical experimentation is required to
determine a good set of cooling parameters for a new problem.
In the work presented here, we have chosen an exponential cooling
schedule and a number of thermalization steps that yields a confidence level
of 95 per-cent for finding the global minimum.
This confidence level was determined by compiling statistics from many
simulations involving 18 randomly distributed quarks.
In each case, we compared configurations arrived at by an exhaustive
search with those given by the annealing algorithm to see if the global
solution was found.

Figure \ref{3qanneal} gives an example of how the simulated annealing
simulation progresses on average for $N=24$ and a low density $\rho=0.05$.
This is in the hadronic phase of the system. The average is over many
Monte Carlo generated instances of the quark coordinates (quark coordinates
are generated according to a variational wavefunction that will be described
in Sec.~\ref{wavefun}). The temperature, acceptance ratio, average potential
energy, and best potential thus far are plotted as a function of the cooling
step. A typical plot of an individual annealing process would show an abrupt
drop in the energy (freezing) somewhere between steps 40 and 55, a common
feature of simulated annealing runs.

\subsubsection{Pairing Algorithms}
\label{pa}

One way to get around the grouping problem described above is to
consider an alternative, albeit simpler, grouping scheme. In this
scheme one considers, instead, pairing the quarks two colors at a time.
Economists have long been interested in the problem of pairing $N$ factories
with $N$ retail stores in order to minimize the overall cost of exchanging
goods. Efficient pairing algorithms with a computational cost proportional
to $N^3$ have already been developed by mathematicians~\cite{burkard80}.
Adapting these efficient algorithms to the problem of interest implies
calling the pairing algorithm three times in succession in order to
independently pair red and green quarks, green and blue quarks, and blue and
red quarks. Notice, however, that the many-quark dynamics must be modified
in order to take advantage of the pairing algorithm. The potential energy
for the system is now given by,
\begin{equation}
 {\bf V}=
  {\rm Min}\sum_{i=1}^{A}\frac{1}{2}(x_{i}^R-x_{i}^G)^2 +
  {\rm Min}\sum_{i=1}^{A}\frac{1}{2}(x_{i}^G-x_{i}^B)^2 +
  {\rm Min}\sum_{i=1}^{A}\frac{1}{2}(x_{i}^B-x_{i}^R)^2 \:,
 \label{potpair}
\end{equation}
and should be compared to Eq.~(\ref{pot}). This version of the
string-flip model has already been used in Ref.~\cite{j1}
and we include some results here for comparison.

In spite of this change both models share many common features.
Clearly, they are identical in the case of an isolated cluster.
Moreover, both guarantee cluster separability thus avoiding
the emergence of long-range van der Waals forces. There are, however,
some differences. Most notoriously, by pairing quarks independently
there is no guarantee that the quarks will be grouped into three-quark
clusters. Color-neutral hadrons in this model may contain any multiple of
three quarks. Thus, the space of string configurations over which the
potential is minimized consists of all colorless clusters containing multiples
of three quarks. We refer to this model as the multi-quark-cluster (MQC) model.
In contrast, the model based on the potential given by Eq.~(\ref{pot}),
which allows for only three-quark clusters, will be referred as the
three-quark-cluster (TQC) model. Since the minimization space is larger in
the MQC model than if only 3-quark clusters were allowed, one can generally
expect solutions which are lower in energy. However, this will not be the
case in the low-density regime, where hadrons are well separated and pairing
quarks independently should, nonetheless, generate only three-quark hadrons.

\section{The Variational Wavefunction}
\label{wavefun}

In this work we use a simple single-parameter variational wavefunction
to represent the ground state of the many-quark system,
\begin{equation}
\Psi=e^{-\lambda\bf V}\Psi_{FG}\:.
\end{equation}
The Fermi-gas wavefunction, $\Psi_{FG}$, is exact for a system of free
fermions with no correlations other than those generated by the Pauli exclusion
principle. It is a product of red, green, and blue Slater determinants.
The exponential factor expresses the degree of clustering through the
variational parameter $\lambda$ with $\lambda^{-1/2}$ being the
length-scale for quark confinement. In spite of the simplicity of the
wavefunction, it is exact, both, in the low-density nuclear matter
phase where the isolated-hadron limit is recovered
($\lambda={1/\sqrt{3}}$) and in the high-density Fermi-gas phase where
$\lambda=0$. The wavefunction regards red, green, and blue quarks as
distinguishable particles (in the same sense as spin-up and spin-down
electrons can be regarded as distinguishable in the presence of a
spin-independent Coulomb interaction). Thus the Pauli principle is enforced
only among quarks of the same color.

\subsection{Determination of $\lambda$}

The ground state expectation values for the kinetic and potential terms of
our Hamiltonian can be related using integration by parts.
The result is,
\begin{equation}
\langle\Psi|T|\Psi\rangle=T_{\rm FG}+
3\lambda^2\langle\Psi|{\bf V}|\Psi\rangle\:.
\end{equation}
$T_{\rm FG}$ is the kinetic energy of a free Fermi gas.
The term $3\lambda^2\langle
{\bf V}\rangle$ is the increase in kinetic energy above the Fermi-gas value
generated by clustering correlations.
Thus, to calculate the total energy we only
need the expectation value of the potential,
\begin{equation}
\langle\Psi|{\bf H}|\Psi\rangle=E(\lambda)=T_{\rm FG}+(3\lambda^2+1)
\langle\Psi|{\bf V}|\Psi\rangle\:.
\end{equation}

As in any variational approach, the variational parameter is determined by
minimizing the energy. A `brute force' implementation is to simply calculate
$E(\lambda)$ for several values of $\lambda$ and use this information to
estimate $\lambda_{\rm min}$. This is very expensive computationally since
several Monte Carlo estimates of $E(\lambda)$ are required for each estimate
of $\lambda_{\rm min}$.

Fortunately, there is a more elegant and efficient method for extracting
$\lambda_{\rm min}$ which takes maximum advantage of every Monte Carlo run.
It makes use of scaling relations for the potential when it is evaluated at
two different values of $\lambda$ and the quark density $\rho$. The condition
for finding $\lambda_{\rm min}$ is the usual vanishing of the derivative of
the energy with respect to the variational parameter
\begin{equation}
\frac{dE}{d\lambda}=6\lambda\langle{\bf V}\rangle-[6\lambda^2+2]
[\langle{\bf V}^2\rangle-\langle{\bf V}\rangle^2]=0\:,
\end{equation}
which requires, in addition to the potential, an estimate of the expectation
value of the square of the potential.

At first glance, $\langle{\bf V}\rangle$ appears to be a function of both
$\lambda$ and $\rho$. However, on the grounds of simple dimensional analysis
one can show that the potential energy can be written (for a fixed number
of quarks) as
\begin{equation}
 \langle{\bf V}\rangle_{\lambda\rho}/L^2=v(\lambda L^2) \:,
\end{equation}
where $\rho=N/L^3$ and $v$ is a dimensionless function of the
scaling variable $\lambda L^2$. Hence, we see that
$\langle{\bf V}\rangle_{\lambda\rho}/L^2$ is in reality a function of
only one scaling variable leading to the following scaling relation for
the potential:
\begin{equation}
\langle{\bf V}\rangle_{\lambda\rho}=(\frac{\rho'}{\rho})^{\frac{2}{3}}
\langle{\bf V}\rangle_{\lambda'\rho'}\:,
\end{equation}
with
\begin{equation}
\lambda'=(\frac{\rho'}{\rho})^{\frac{2}{3}}\lambda\:.
\end{equation}

We can use this freedom as follows.
After performing a calculation using $\rho,\lambda$ we can always
rescale our results to a new pair of variables, $\rho',\lambda'$, such that
$dE/d\lambda=0$ is satisfied. After some algebra, we find the new
density which accomplishes this is given by,
\begin{equation}
\rho'=\rho[3\lambda(\frac{\langle{\bf V}\rangle_{\lambda\rho}}
{\langle{\bf V}^2\rangle_{\lambda\rho}-
\langle{\bf V}\rangle^2_{\lambda\rho}}-\lambda)]^{-\frac{3}{4}}\:.
\end{equation}
This is indeed a useful result since it now takes only one Monte Carlo
estimate of $\langle{\bf V}\rangle$ and $\langle{\bf V}^2\rangle$ to determine
the variational parameter. The only potential pitfall of this method
for acquiring $\lambda_{\rm min}$ is that $E(\lambda)$ should be a smooth
function without multiple minima. One should therefore be careful in using this
technique, particularly near the transition between the confined and unconfined
phases of the system, where double minima or degeneracies might occur.

\section{Simulation Results}
\label{res}

Figure \ref{3qdiff} shows the difference in the potential energies
in the MQC model (obtained by using the pairing algorithm) and the
TQC model (with simulated annealing) as a function of the density
of the system. As expected, the two agree at very low density where
the hadrons are well separated and multi-quark configurations are
suppressed. At a higher density, however, the MQC model takes
advantage of the possibility of forming large multi-quark rings
and yields a lower energy.

Figure \ref{3qenergy3} shows the energy per quark as a function of density
for the MQC model ($N=24,96$) and for the TQC model ($N=24$) along with
the Hartree-Fock ($\lambda=0$) result.
Deviations from the Hartree-Fock result in the high-density phase are
indicative of finite-size effects present in the simulation and get smaller
as $N$ is increased in the MQC model as one might expect.
While this finite-size effect lowers the energy per quark
in the quark-matter phase, it has little effect in the nuclear-matter phase.
The transition between the two phases occurs abruptly near $\rho=0.12$.

Figure \ref{3qenergy3} also reveals that the TQC model with $N=24$
shows roughly the same finite-size effect as the $N=96$ MQC result.
One might expect larger finite-size effects for the latter since the MQC
model allows the formation of large multi-quark clusters which can
have dimensions comparable to the size of the box. In contrast, we have found
that three-quark clusters are always considerable smaller than the size
of the box. Based on this evidence, we will compare $N=24$ annealing results
with $N=96$ pairing results since they seem to have comparable finite size
effects.

The variational parameter $\lambda$ as a function of density is given in
Figs.~\ref{3qlvd1},~\ref{3qlvd2},~\ref{3qlvd3}.
These figures show the transition
between the hadronic phase where $\lambda$ is near the isolated hadron value,
and the quark-gas phase characterized by $\lambda$ tending to zero.
The MQC model shows a step-like behavior in $\lambda(\rho)$ at the
transition density, whereas the TQC model gives a gradual, continuous
transition. This can be understood in terms of the structure of
$E(\lambda)$ near the phase transition. $E(\lambda)$ generated by the MQC
model is characterized by the formation of double minima near the transition
giving rise to an abrupt shift to a smaller value of $\lambda$~\cite{j1} .
In contrast, we have found that in the annealing model, $E(\lambda)$ becomes
very flat (constant) near the phase transition, resulting in large
fluctuations in the string lengths and a corresponding broad and smooth
transition between the two phases (the nature of the transition for a
larger number of quarks in the TQC model remains to be investigated).

The phase transition is demonstrated most dramatically in Fig.~\ref{3qenergy}
where the abrupt increase in the potential energy (increasing string length)
and a rapid decrease in kinetic energy (lower Fermi energy, deconfinement) can
be seen at the crossover point. At very low density the equipartition of energy
between kinetic and potential forms is satisfied, and at high density, the
system is dominated by the Fermi-driven kinetic component as expected from the
behavior of a free quark gas. In Figure \ref{3qavgr} the average string
length was directly measured as a function of density. The length is presented
in units of the isolated cluster value which can be calculated exactly.
Again, we see the abrupt swelling of the clusters near the transition density.
As expected, the behavior of the average string length is identical to that
of the potential energy in Fig.~\ref{3qenergy}.

\section{Conclusion}
\label{concl}

In this paper, we have presented the main results of a three-quark string-flip
model using simulated annealing to determine optimal groupings. Previous
work with a similar model used pairing to achieve the grouping.
A major difference between the models is the presence of multi-quark clusters
in the MQC model, while the TQC model allows for only 3-quark hadrons.

To summarize, both models have the expected low density (isolated hadrons) and
high density (quark-gas) behavior. Also, the transition density is roughly the
same in both models. The main differences are that the TQC model appears
to have smaller finite size effects due to the 3-quark constraint, and
the variational parameter, $\lambda(\rho)$, is smooth and continuous across
the phase transition whereas, in the MQC model, $\lambda(\rho)$ is
step-like and discontinuous.

The immediate followup to this work will be to use the values determined
here for the variational parameter, $\lambda$, to calculate additional
ground-state observables like the quark-quark correlation function.
This might further characterize the hadron to quark-gas transition and
could be important in calculating `quark giant resonances'. Longer range
projects may include a spin-dependent interaction (which is responsible
for the $N-\Delta$ mass splitting) and light-quark flavor degeneracy in
the hope of understanding the saturation of nuclear matter using only
quark degrees of freedom. In addition, one might include additional
(heavy) flavors to study and characterize the transition to strange matter.

\acknowledgments
This research was supported by the Florida State University
Supercomputer Computations Research Institute and U.S. Department
of Energy contracts DE-FC05-85ER250000, DE-FG05-92ER40750.

\figure{$\Delta$ and Y coupling schemes for quarks. For harmonic potentials,
         they are equivalent.
         \label{3q01}}
\figure{Annealing algorithm used for solving the string assignment
        problem.
        \label{3qalg}}
\figure{Various quantities involved in the annealing process as a function
        of the number of cooling steps. This figure is an average result
        over many Metropolis steps for 24 quarks.
        \label{3qanneal}}
\figure{Percent difference between optimal potential energies as returned by
        the TQC model (annealing algorithm) and the MQC model (pairing
        algorithm). The larger
        configuration space of the pairing process yields lower energies
        except at very low density where the difference tends to zero.
        \label{3qdiff}}
\figure{Comparison of the total energy per quark in the TQC model vs. the
        MQC model for 24 and 96 quarks. Also shown is the Hartree-Fock
        ($\lambda=0$) result.
        \label{3qenergy3}}
\figure{Variational parameter vs. density from the MQC model with $N=24$
quarks.
        \label{3qlvd1}}
\figure{Variational parameter vs. density from the MQC model with $N=96$
quarks.
        \label{3qlvd2}}
\figure{Variational parameter vs. density from the TQC model with $N=24$
quarks.
        \label{3qlvd3}}
\figure{Results of the TQC model using simulated annealing for 24 quarks.
        Total, kinetic and potential energy per quark are shown together with
        the Hartree-Fock ($\lambda=0$) result. The vertical dashed line
        indicates the transition region between the hadronic and quark phases.
        \label{3qenergy}}
\figure{Average string length (units of isolated hadron value) as a function
        of density in the TQC model.
        \label{3qavgr}}

\clearpage


\begin{references}
\bibitem{j2} F. Lenz {\it et al.}, Ann. of Phys. {\bf 170} (1986) 65.
\bibitem{h1} C. J. Horowitz, E. J. Moniz, and J. W. Negele, Phys. Rev.
             D {\bf 31}, 1689 (1985).
\bibitem{j3} C. J. Horowitz and J. Piekarewicz, Phys. Rev. C {\bf 44},
             2753 (1991).
\bibitem{j1} C. J. Horowitz and J. Piekarewicz,
             Nucl. Phys. {\bf A536}, 669 (1992).
\bibitem{Otten} R. H. J. M. Otten and L. P. P. P. van Ginneken,
                {\em The Annealing Algorithm} (Kluwer Academic Publishers,
                Boston, 1989).
\bibitem{burkard80} R.E.~Burkard and U.~Derigs, {\it Lecture Notes in
                    Economics and Mathematical Systems} (Springer-Verlag,
                    Berlin, 1980), Vol.~184.
\end{references}
\end{document}